\documentclass{sigchi-ext}
\usepackage[T1]{fontenc}
\usepackage{textcomp}
\usepackage[scaled=.92]{helvet} 
\usepackage{graphicx} 
\usepackage{balance}  
\usepackage{booktabs} 
\usepackage{ccicons}  
\usepackage{ragged2e} 

\def\plaintitle{Categorizing Sources of Information for Explanations in Conversational AI Systems in the Home for Older Adults Aging in Place} 
\def\emptyauthor{}
\def\plainkeywords{Authors' choice; of terms; separated; by
  semicolons; include commas, within terms only; required.}

\title{Categorizing Sources of Information for Explanations in Conversational AI Systems for Older Adults Aging in Place}

\numberofauthors{3}

\author{%
  \alignauthor{%
    \textbf{Niharika Mathur}\\
    \affaddr{Georgia Institute of Technology} \\
    \affaddr{Atlanta, GA, USA} \\
    \email{nmathur35@gatech.edu} }\alignauthor{%
    \textbf{Elizabeth Mynatt}\\
    \affaddr{Northeastern University}\\
    \affaddr{Boston, MA, USA}\\
    \email{e.mynatt@northeastern.edu} } \vfil \alignauthor{%
    \textbf{Tamara Zubatiy}\\
    \affaddr{Georgia Institute of Technology}\\
    \affaddr{Atlanta, GA, USA}\\
    \email{tzubatiy3@gatech.edu} }}

\definecolor{linkColor}{RGB}{6,125,233}
\hypersetup{%
  pdftitle={\plaintitle},
  pdfauthor={\emptyauthor},
  pdfkeywords={\plainkeywords},
  bookmarksnumbered,
  pdfstartview={FitH},
  colorlinks,
  citecolor=black,
  filecolor=black,
  linkcolor=black,
  urlcolor=linkColor,
  breaklinks=true,
}

\begin{document}

\CopyrightYear{2024}
\setcopyright{rightsretained}

\copyrightinfo{\textit{Permission to make digital or hard copies of part or all of this work for personal or
classroom use is granted without fee provided that copies are not made or distributed
for profit or commercial advantage and that copies bear this notice and the full citation
on the first page. Copyrights for third-party components of this work must be honored.
For all other uses, contact the owner/author(s).
Copyright held by the owner/author(s).
ACM CHI’24, May 11-16, 2024, Honolulu, HI, USA}}

\maketitle

\RaggedRight{} 

\begin{abstract}
 As the permeability of AI systems in interpersonal domains like the home expands, their technical capabilities of generating explanations are required to be aligned with user expectations for transparency and reasoning. This paper presents insights from our ongoing work in understanding the effectiveness of explanations in Conversational AI systems for older adults aging in place and their family caregivers. We argue that in collaborative and multi-user environments like the home, AI systems will make recommendations based on a host of information sources to generate explanations. These sources may be more or less salient based on user mental models of the system and the specific task. We highlight the need for cross technological collaboration between AI systems and other available sources of information in the home to generate multiple explanations for a single user query. Through example scenarios in a caregiving home setting, this paper provides an initial framework for categorizing these sources and informing a potential design space for AI explanations surrounding everyday tasks in the home.

\end{abstract}

\section{Introduction}
The vision of Human-Centered Explainable AI (HCXAI) emphasizes that merely providing an explanation for an AI decision is insufficient; it must be situated sociotechnically and contextually within the user's environment to be effective \cite{ehsan2023charting}. In this paper, we align our work with the workshop’s research objectives by focusing on the following question: \textit{Whose “voices” are represented in AI explanations?}. In proceedings of the Workshop on Human-centered Explainable AI at CHI (2022), Alizadeh et al highlight limitations of commercial Conversational AI systems (CAIs) in facilitating an effective conversation with users leading to user confusion and frustration \cite{hcxai2022}. These limitations also increase user burden in employing recovery strategies to establish a somewhat successful interaction \cite{beneteau2019communication, sciuto2018hey, zubatiy2021empowering}. Authors in \cite{hcxai2022} further argue that explanations in AI systems need to consider a “holistic user experience” and not address single incident queries. Building on this recommendation, we argue that this holistic understanding of the user is crucial in generating contextual AI explanations in CAIs. Our aim is to explore this holistic understanding of the user in further detail, grounded in our work to support older adults aging in place.  CAIs (smart speakers or chatbots) are a potent testing ground for explanations since they encapsulate explainability as a dialog problem between the AI agent and the user, and can be used to design Wizard of Oz experiments to evaluate different explanatory conditions \cite{jentzsch2019conversational}. Such AI systems are also equipped with the capability to store and recall user interactions over time, thereby providing an opportunity to explore the potential of an enhanced holistic understanding of the user context. 

In this paper, our primary argument is that CAIs operating within collaborative and multi-user settings like the home must leverage diverse information sources to build affordances that can generate explanations tailored to user motivations. Consider a scenario when an AI system reminds an older adult (John) to take his medication. John, in return, asks why the reminder was made at the time. In the case that John is motivated by monitoring physical items associated with the medication task such as his pillbox, the AI should have the ability to sense information that his pillbox has not been opened in a few hours, and hence use that to explain the timing of the reminder. Alternatively, if John is motivated by focusing on routines adjacent to the medication task, the AI should be able to explain that since John was late in getting home this evening, he likely may have missed his medication with dinnertime and provide additional data to explain and contextualize the reminder. 

Given the varied sources of information available to AI systems, there's potential to generate explanations of differing saliency levels. Our aim is to explore explanations that instill the most confidence in users, considering their mental models and tasks. In this workshop, we hope to discuss an initial framework categorizing these information sources and the challenges of designing effective explanations based on diverse mental models and user preferences, particularly focusing on older adults with MCI. Unlike subject matter experts or younger demographics, older adults are unlikely to have a detailed mental model of AI systems. Given this, possible AI explanations will need to draw from characteristics that are salient in their daily lives, such as routines, preferences and data from previous interactions and common household items. 

\section{Contribution}
Efforts have been made to categorize information for explanations \cite{arya2019one, wang2019designing, guidotti2018survey}, yet they lack specificity regarding contextual understanding from home environments. Liao et al \cite{liao2021question} focus on explaining information and designing suitable presentation formats, emphasizing iterative design and evaluation. We argue that AI systems should integrate diverse information sources to bridge the sociotechnical gulf between expectation and experience \cite{luger2016like}. To support this, we propose an initial framework categorizing home information sources and their importance in explanation design. We offer examples for each category to aid AI designers and engineers, suggesting ways to integrate data streams from these sources into AI explanations. These examples stem from ongoing research on interactions among older adults, caregivers, and AI systems. Our aim is to foster discussions on user-centric explanations by considering multiple external sources of information beyond AI algorithms.

\section{Method}
 
The framework in this paper is grounded in findings from 2 longitudinal studies of 20 weeks and 10 weeks involving 7 and 10 pairs respectively of older adults with MCI and their caregivers. The aim of these studies was to analyze the longitudinal usage of a CAI in the home, specifically for medication management in \cite{mathur2022collaborative}. Participants interacted with the CAI daily over the study duration, facilitated by a conversational medication check-in routine in \cite{mathur2022collaborative}. Our research team analyzed 1596 total initiated interactions from the AI system and we recorded 844 responses from older adults and analyzed them manually for patterns and themes. We also conducted semi-structured interviews at different points in the studies to delve deeper into participant experiences and challenges with the AI system. Further details about these studies can be found in our full papers that discuss the design, deployment and evaluation of the longitudinal usage of the AI and of the medication assistant in depth with the participants \cite{mathur2022collaborative, zubatiy2021empowering}.

\section{Framing the Context}
Our framework for categorizing information sources serves two main purposes. Firstly, it acts as a foundation for navigating user data within the home environment, shedding light on the variety of accessible information for AI systems. Secondly, we aim to underscore the algorithm-centric nature of AI explanations as a checklist item rather than a human-centric feature. This perspective aligns with the emphasis on understanding the "who" of explanations, as advocated by Ehsan et al. \cite{ehsan2022human}. Our position emphasizes the personalized nature of explanation, where each user may necessitate tailored explanations from different sources based on their unique context. Therefore, our framework advocates for referencing diverse knowledge sources and promoting cross-technological collaboration within the home to generate multiple explanations, with the selection depending on the specific context of the user query.

\begin{figure*}
  \includegraphics[width=2\columnwidth]{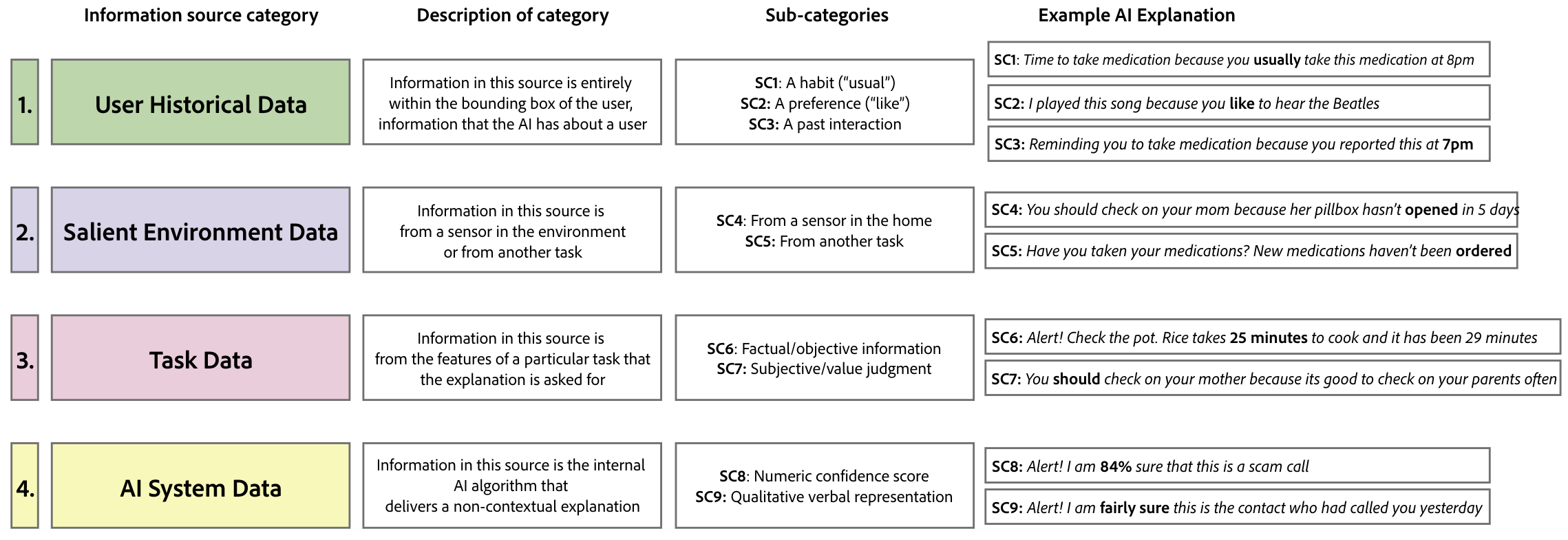}
  \caption{This initial framework categorizes sources of information in the home into 4 larger categories with 9 subcategories.}
\end{figure*}

\section{The Initial Framework}

We first specify a larger information category, and then subcategories within it. Each sub-category has an example explanation in Figure 1 drawn from scenarios in assisted family caregiving for older adults. The subcategories within the framework are not mutually exclusive; rather, they illustrate how an explanation can be represented differently through various information sources, depending on what a user is motivated by.

\subsection{\textbf{Category 1 - User Historical Data}} 

Source of information in this category is the \textbf{user}. This includes information that was conveyed to the AI in historical interactions either \textbf{\textit{by}} a user; or \textbf{\textit{for}} a user. In both instances, the AI stores the information as an attached piece of knowledge about the user to append in an explanation. There are three ways in which information can be represented in an explanation in this category - \textit{\textbf{As a habit (“usual”)}} or \textit{\textbf{As a preference (“likeness”)}} or \textit{\textbf{A definite past interaction.}} 

\subsection{\textbf{Category 2 - Salient Environment Data}} 

Source of information in this category is information that the AI has gleaned either from an environmental sensor or from another task, i.e., information acquisition happens through a source \textit{\textbf{external}} to the user’s bounding box. There are two ways in which information can be represented in an explanation in this category - \textit{\textbf{From a smart sensor in the home,}} or \textit{\textbf{From another task (task that is different from the original task that the explanation is sought for)}}

\subsection{\textbf{Category 3 - Task Data}} 

Source of information in this category is the contextual \textbf{specifics} or \textbf{features} of the task that the explanation is sought for. This is more relevant in complex tasks such as cooking or medication. There are two ways in which information can be represented in an explanation - \textit{\textbf{Explanation with a factual/objective information}}, or \textit{\textbf{Explanation with a subjective value/moral judgment}}  

\subsection{\textbf{Category 4 - AI System Data}}   

Source of information in this category is the \textbf{internal AI algorithm} that delivers a non-contextual explanation without information from a past interaction, a user preference, or a sensor data. This can be thought of as a last resort for the AI to provide an explanation when either the other sources do not provide significant data or user action is not motivated by the other sources. There are two ways in which information can be represented in an explanation in this category - \textit{\textbf{Explanation with a numeric confidence score}}, or \textit{\textbf{Explanation with a qualitative verbal representation}}  

\section{Challenges and Technical Opportunities}

While we are excited about the potential of the presented initial framework to generate user-tailored explanations, there are a few technical opportunities and challenges that we would like to address concerning the implementation of this framework. Among other potential technical systems, we currently envision this framework to be implemented in a feedback loop model that begins with delivering a default explanation to the user at the start, then takes in the user feedback and models that to generate a new explanation from the framework, thereby increasing its user understanding in the process (somewhat similar to a “learning from demonstration” model in robotics). In this model, we envision a potential challenge in being able to capture user feedback that is not explicit in nature, i.e., developing a system that captures implicit feedback in the form of open-ended questions, or a survey, and models that or learns from it to deliver a new explanation. Another potential challenge is that of user privacy when extrapolating data coming in from home sensors to build into explanations. This would require the implementation of security frameworks for sensitive user data coming in from sensors, as well as the incorporation of individual user attitudes towards privacy conscientiousness.        
\section{Conclusion and Future Work}

In this paper, we introduce an initial framework for categorizing home information sources to enhance AI explanations for older adults aging in place for everyday tasks. We advocate for diverse explanations drawn from various user-related information such as their routines, preferences, and activities. Future work will assess the effectiveness of these explanations for older adults with varying motivations and expectations, additionally evaluating their impact on user confidence based on the conveyed information types and their affordances.

\balance{} 

\bibliographystyle{SIGCHI-Reference-Format}
\bibliography{bibliography}

\end{document}